# Towards Antihydrogen Confinement with the ALPHA Antihydrogen Trap[*]

M. C. Fujiwara

*TRIUMF, Vancouver, British Columbia, Canada*
Makoto.Fujiwara@triumf.ca

G. Andresen

*University of Aarhus, Aarhus, Denmark*

W. Bertsche

*University of California at Berkeley, Berkeley, California, USA*

A. Boston

*University of Liverpool, Liverpool, UK*

P. D. Bowe

*University of Aarhus, Aarhus, Denmark*

C. L. Cesar

*Federal University of Rio de Janeiro, Rio de Janeiro, Brazil*

S. Chapman

*University of California at Berkeley, Berkeley, California, USA*

M. Charlton

*University of Wales Swansea, Swansea, UK*

M. Chartier

*University of Liverpool, Liverpool, UK*

A. Deutsch

*University of California at Berkeley, Berkeley, California, USA*

J. Fajans

*University of California at Berkeley, Berkeley, California, USA*

R. Funakoshi

---

[*]Invited talk at International Conference on Trapped Charged Particles and Fundamental Physics: TCP06, Parksville, BC, Canada, September 2006.

*University of Tokyo, Tokyo, Japan*

D. R. Gill

*TRIUMF, Vancouver, British Columbia, Canada*

K. Gomberoff

*Technion, Haifa, Israel*

J. S. Hangst

*University of Aarhus, Aarhus, Denmark*

W. N. Hardy

*University of British Columbia, Vancouver, British Columbia, Canada*

R. S. Hayano

*University of Tokyo, Tokyo, Japan*

R. Hydomako

*University of Calgary, Calgary, Alberta, Canada*

M. J. Jenkins

*University of Wales Swansea, Swansea, UK*

L. V. Jørgensen

*University of Wales Swansea, Swansea, UK*

L. Kurchaninov

*TRIUMF, Vancouver, British Columbia, Canada*

N. Madsen

*University of Wales Swansea, Swansea, UK*

P. Nolan

*University of Liverpool, Liverpool, UK*

K. Olchanski

*TRIUMF, Vancouver, British Columbia, Canada*

A. Olin

*TRIUMF, Vancouver, British Columbia, Canada*

R.D. Page, Liverpool

*University of Liverpool, Liverpool, UK*

A. Povilus

*University of California at Berkeley, Berkeley, California, USA*

F. Robicheaux

*Auburn University, Auburn, Alabama, USA*

E. Sarid

*Nuclear Research Center Negev, Beer Sheva, Israel*

D. M. Silveira

*Federal University of Rio de Janeiro, Rio de Janeiro, Brazil*

J. W. Storey

*TRIUMF, Vancouver, British Columbia, Canada*

R. I. Thompson

*University of Calgary, Calgary, Alberta, Canada*

D. P. van der Werf

*University of Wales Swansea, Swansea, UK*

J. S. Wurtele

*University of California at Berkeley, Berkeley, California, USA*

Y. Yamazaki

*RIKEN, Saitama, Japan*

(ALPHA Collaboration)

ALPHA is an international project that has recently begun experimentation at CERN's Antiproton Decelerator (AD) facility. The primary goal of ALPHA is stable trapping of cold antihydrogen atoms with the ultimate goal of precise spectroscopic comparisons with hydrogen. We discuss the status of the ALPHA project and the prospects for antihydrogen trapping.

## Introduction

Antihydrogen, an atomic bound state of an antiproton and a positron, represents the simplest system of atomic antimatter, and is an ideal system for testing symmetries between matter and antimatter. The first atoms of antihydrogen were observed at relativistic velocities [1, 2], such that precision measurements of its properties were not possible. Cold antihydrogen atoms were first produced and detected in the ATHENA experiment in 2002 [3]. This was followed by a similar observation by the ATRAP collaboration [4]. Since then, rapid progress has been made in this emerging field [5]. A next major goal is stable trapping of cold antihydrogen atoms. In order to probe matter-antimatter symmetry with the highest possible precision using antihydrogen, it is essential that the anti-atoms be suspended in vacuum to allow detailed studies. The new ALPHA apparatus was designed and built to do just that. In this paper, we give a general overview of the ALPHA experiment, together with some of the results obtained during the collaboration's first AD runs in 2006.[†]

## Motivation for trapped antihydrogen

One of the prime physics motivations for antihydrogen studies is tests of CPT invariance. The combined operation of Charge-conjugation, Parity, and Time reversal is believed to be an exact symmetry of Nature, due to the existence of the CPT theorem [6]. However, the assumptions of this theorem may not be valid at some very high energy scale, or equivalently at very short distances. For example at the Planck scale, where the gravitational interaction becomes important, possibilities exist for CPT violation [7, 8]. Therefore, achieving experimental sensitivities necessary to probe Planck scale physics is a benchmark for any CPT test.

So where is the best place to look, to see violation of CPT invariance? Given, unfortunately, that there exist no accepted fundamental theories predicting the magnitude and the pattern of CPT violation, one may consider two approaches; either strive to achieve the highest possible energies, or the highest possible precision. On the high energy front, CPT and Lorentz invariance violations are searched for, e.g., in high energy cosmic rays [7, 8]. Clearly, the spectral comparison of hydrogen and antihydrogen belongs to the precision category, together with other existing tests with elementary particles such as neutral K mesons [9]. Properties of atomic hydrogen, such as the 1s-2s optical transition frequency, and the ground state hyperfine splitting, are some of the most precisely measured quantities in nature. Hence they offer challenging targets for antihydrogen measurements, with which, ultimately, Planck scale sensitivity could be expected [10, 11]. It is worth noting that CPT violation may be responsible for the cosmological asymmetry between matter and antimatter [12-15].

---

[†] Some of the results presented here were obtained after the TCP06 conference, but are included for completeness.

## ALPHA Overview

ALPHA is an ambitious project, with a demanding schedule, in which it has been necessary to construct much of the apparatus from scratch. The exception is the positron accumulator, which was inherited from ATHENA [16]. Tremendous progress has been made since the approval of the project by CERN in June 2005, such that some important results were obtained during the commissioning period from August to November 2006.

ALPHA consists of an ATHENA-type versatile Penning trap, with a superimposed magnetic trap for neutral anti-atom trapping [17]. The latter comprises a magnetic configuration which has a three-dimensional minimum which can trap neutral (anti)atoms via the magnetic interaction $-\vec{\mu} \cdot \vec{B}$, where $\vec{\mu}$ is the magnetic moment of the (anti)atoms and $\vec{B}$ is the magnetic field. If antihydrogen atoms can be created inside this apparatus with low enough speeds, they can be trapped magnetically. The trapping depth, *U,* for ground state antihydrogen is given by:

$$U = 0.7\Delta B \quad (K), \qquad (1)$$

where ΔB is the difference between the minimum and maximum fields in T. This means that it is necessary to produce antihydrogen with kinetic energies equivalent to Kelvin-scale temperatures, since ΔB cannot be made more than a few T with the present superconducting technology (especially for radial confinement).

The magnetic field minimum required for atom trapping can be achieved using a combination of a multipolar field for radial confinement and a pair of coils that provide a mirror field for axial confinement. The trapping field ΔB in Eq. (1) is given, for the axial direction, by $\Delta B_z = B_m$, where $B_m$ is the strength of the mirror field. On the other hand, for the radial direction, it can be written:

$$\Delta B_r = \sqrt{B_s^2 + B_r^2} - B_s, \qquad (2)$$

where $B_s$ is the field strength of a solenoid used for charge particle trapping, and $B_r$ is the radial field strength.

The importance of using a small background field $B_s$ is illustrated in Fig. 1, where the normalized trapping depth is plotted as a function of the axial background field, for radial fields of 0.6 T, 1.2 T, and 1.8 T. As can be seen, reducing $B_s$ from 3 T to 1 T will result in an increase of the trapping depth by more than factor of 2 for the same radial field. However, if only one solenoid magnet is used, the reduced $B_s$ results in a reduced efficiency for trapping antiprotons. As can be seen in Fig. 2, the antiproton trapping efficiency achieved by ALPHA at 1 T is nearly an order of magnitude smaller than that at 3 T.

In order to solve this dilemma, ALPHA has adopted a two-solenoid approach (Fig. 3), where a high efficiency of antiproton trapping is achieved in a 3 T region, while a lower background field of 1 T is maintained in the trapping region to maximize, as far as is practical, the trapping depth for the neutrals. We have recently shown that antihydrogen can be produced in the 1 T (but radially uniform) background field [18].

## Antimatter plasmas in an octupole trap

A major challenge which needs to be overcome before antihydrogen can be trapped is to reduce the influence of the strongly non-uniform radial magnetic field on the trapped charged particles. Obviously, the antiproton and positron clouds need to be trapped long enough for antihydrogen to be formed efficiently. Conventional neutral atom traps achieve radial confinement using quadropolar magnetic fields. However, the stability of trapped charge particles in the presence of those fields has been the subject of controversy. In recent measurements, Fajans et al. have reported a large loss of charged particles in the presence of a strong quadropolar field, and identified an important particle loss mechanism partly responsible for this instability, known as ballistic loss [19]. In this process, particles that have radii greater than a certain critical radius $r_c$ (see [19]) follow magnetic field lines which intercept the trap walls, and are lost immediately. In ALPHA, we have adopted an octupolar field configuration to alleviate these losses. Fig. 4 shows a comparison of the radial field profile for ideal quadropole and octupole coil geometries and illustrates how the perturbation by the field non-uniformity near the trap axis, where particles are initially stored, is much reduced for the octupole case. Recently we have shown that charged particles can indeed survive for long periods in an octupolar field, as illustrated in Figs. 5 and 6 [20]. In these measurements, we have separately stored positrons and antiprotons in an octupole field of strength $B_r$=1.2 T in a background field of $B_s$=1 T, which would result in a trapping potential equivalent to 0.4 K for ground state antihydrogen. In another series of measurements, interactions between antiprotons and positrons in a nested Penning trap configuration have been observed. Furthermore first attempts were made to trap antihydrogen [21]. These results from our commissioning run constitute major milestones towards antihydrogen trapping.

## ALPHA antihydrogen detector

Currently a new imaging detector for detection of antihydrogen is being constructed to facilitate the unambiguous identification of trapped antihydrogen. Antihydrogen annihilation events will be detected by reconstructing the trajectories of the charged particles (mostly pions) produced by antiproton annihilations via position sensitive silicon strip detectors. A similar detector has been used for imaging antiproton annihilations in ATHENA [22], but the conditions imposed by the requirements for the neutral trap are more stringent. In order to achieve the maximum trapping depth for the neutrals, we must have the octupole magnet as close as possible to the trap electrode wall. This necessitates that the detector be placed outside the superconducting magnet and the cryostat. The challenges in this configuration include increased multiple scattering of the pions, and a large distance between the annihilation point and the first layer of the detector, both resulting in a degraded position resolution. Detailed studies show that these challenges are manageable by having sufficiently fine pitches in the silicon strip detectors [23]. In addition, the increased trap size of ALPHA, about a factor of two larger in radius than used by ATHENA, requires a larger detector area. Figure 7 illustrates expected annihilation images with the ALPHA detector from detailed GEANT Monte Carlo simulations, which include a realistic geometry, antiproton annihilation branching ratios, and a vertex reconstruction algorithm based on helix fitting to charged tracks in the magnetic field [23].

## Physics with trapped antihydrogen

Once antihydrogen is successfully trapped, a number of possibilities for new measurements would become available. One of the most promising measurements is 1s-2s two photon spectroscopy, which is a prime goal of ALPHA. Here, let us discuss another possibility, ground state hyperfine spectroscopy in a high magnetic field. Note that a hyperfine measurement in a low field has been proposed with an atomic beam method [24].

The energy levels of (anti)hydrogen atoms in a magnetic field, given by the Breit-Rabi formula, are shown in Fig. 8. Microwave transitions from low field seeking, trappable states (c, d in Fig. 8) to high field seeking un-trappable states (a, b) would lead to the ejection of the antihydrogen from the trap. The transition can be detected, with nearly 100% efficiency, via annihilation signals on the trap wall measured by ALPHA's position sensitive vertex detector. An onset of this signal is expected to occur when the frequency of an applied microwave is varied such that the anti-atoms come into resonance. Each transition frequency is sensitive to a different combination of fundamental parameters. For example, in the high-field limit the transitions between the states (d→ a) and (c→b) are given by:

$$\nu_{ad} = \frac{\Delta\nu}{2} + \frac{\mu_B g_{e+}}{h} B$$
$$\nu_{bc} = -\frac{\Delta\nu}{2} + \frac{\mu_B g_{e+}}{h} B$$

(3)

where $\Delta\nu$ is the antihydrogen hyperfine splitting at zero field, $\mu_B$ the Bohr magneton, and $g_{e+}$ the positron g-factor in the antihydrogen atom. The difference ($\nu_{ad}$-$\nu_{bc}$) = $\Delta\nu$ will give a measure of the antihydrogen hyperfine splitting, which is directly proportional to the antiproton magnetic moment. The latter is only known with 0.3% accuracy [9]. Therefore a measurement of $\Delta\nu$ with a precision of 0.1% or better would already constitute a significant CPT test of the magnetic properties of an anti-baryon. The sum ($\nu_{ad}$+ $\nu_{bc}$) will give the value of the positron bound state g-factor, which has never been measured due to the lack of stable positronic atoms.

Experimentally, microwave spectroscopy of magnetically trapped neutral atoms has been demonstrated by Pritchard's group [25], but many technical challenges need to be overcome for antihydrogen. These include the higher magnetic field non-uniformity, the presumably higher kinetic energy of the anti-atoms and their as yet unknown quantum state distributions, to name a few. However, demonstrating microwave resonant ejection would already be already a substantial technical achievement, and if the spectroscopy is successful, it will give important information on an anti-atomic system.

## Summary and outlook

We have given an overview of the ALPHA project. Good progress has been made in the first year of ALPHA, and several important milestones have been achieved. A new imaging detector will be installed in 2007. We are hopeful that exciting results will emerge from the ALPHA experiment in the coming years.

### Acknowledgements

We are grateful to the AD team at CERN for providing a high quality antiproton beam. We wish to thank CERN, BNL and TRIUMF for the technical support they have provided. This work was supported by CNPq, FAPERJ, CCMN/UFRJ (Brazil), ISF (Israel), MEXT, RIKEN (Japan), FNU (Denmark), NSERC, NRC/TRIUMF (Canada), DOE, NSF (USA), and EPSRC (UK).

### References

[1] G. Baur et al., Phys. Lett. B **368**, 251 (1996).
[2] G. Blanford et al., Phys. Rev. Lett. **80**, 3037 (1998).
[3] M. Amoretti et al., Nature (London) **419**, 456 (2002).
[4] G. Gabrielse et al., Phys. Rev. Lett. **89**, 213401 (2002).
[5] For a recent review, see e.g. M. Amoretti et al., Nucl. Instr. Meth. B **247**, 133 (2006)
[6] G. Luders, Ann. Phys. **2**, 1 (1957).
[7] See, e.g, V. A. Kostelecky (Ed.), Proc. of 3$^{rd}$ meeting of CPT and Lorentz Symmetry (World Scientific, Singapore, 2004).
[8] M. Pospelov and M. Romalis, Phys. Today, Vol. **57** Num. 7, P. 40 (2004).
[9] Particle Data Group, J. Phys. G **33**, 1 (2006).

[10] R. Bluhm, V. A. Kostelecky, N. Russell, Phys. Rev. Lett. **82**, 2254 (1999).

[11] G. M. Shore, Nucl. Phys. B **717**, 86 (2005).

[12] A. D. Dolgov and Ya. B. Zeldovich, Rev. Mod. Phys. **53**, 1 (1981).

[13] O. Bertolami et al., Phys. Lett. B **395**, 178 (1997).

[14] S. M. Carroll and J. Shu, Phys. Rev. D **73**, 103515 (2006).

[15] P. A. Bolokhov and M. Pospelov, Phys. Rev. D **74**, 123517 (2006).

[16] L. V. Jørgensen et al., Phys. Rev. Lett. **95**, 025002 (2005).

[17] W. Bertsche et al. (ALPHA Collaboration), Nucl. Instr. Meth. A **566**, 746 (2006).

[18] G. Andresen et al., (ALPHA Collaboration), submitted to Phys. Lett. B.

[19] J. Fajans et al., Phys. Rev. Lett. **95**, 155001 (2005).

[20] G. Andresen et al., (ALPHA Collaboration), Phys. Lev. Lett. **98**, 023402 (2007).

[21] G. Andresen et al., (ALPHA Collaboration), in preparation.

[22] M. C. Fujiwara et al., Phys. Rev. Lett. **92**, 065005 (2004).

[23] M. C. Fujiwara, Proc. of Physics with Ultra Slow Antiproton Beams, AIP Conf. Proc. **793**, 111 (2005); M. C. Fujiwara et al., ALPHA Technical Report (2005, unpublished).

[24] E. Widmann et al., Nucl. Instr. Meth. B **214**, 31 (2004).

[25] A. G. Martin et al., Phys. Rev. Lett. **61**, 2431 (1988).

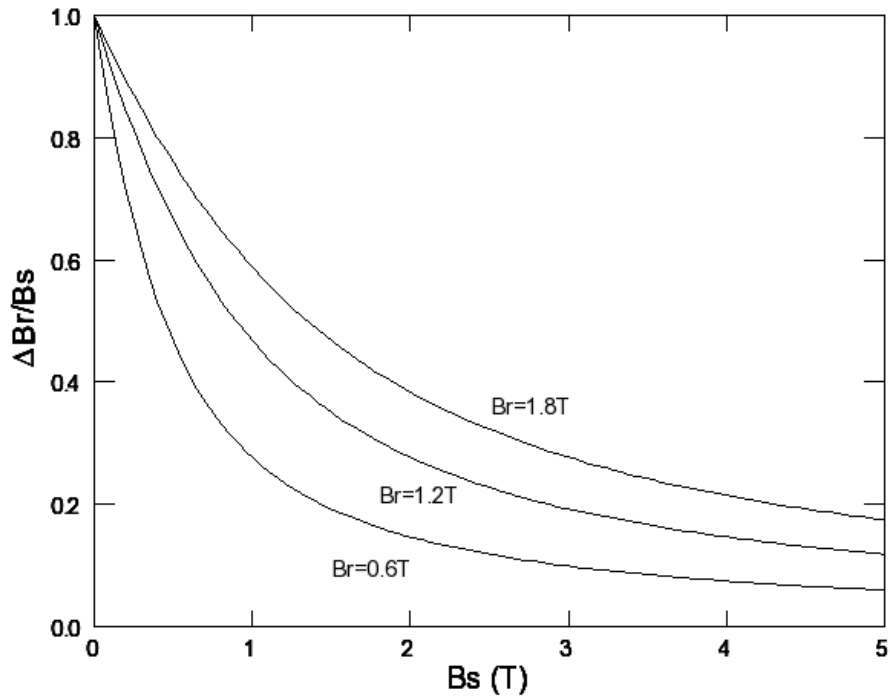

Fig. 1: Normalized trapping depth versus background solenoid field strength for varying radial fields.

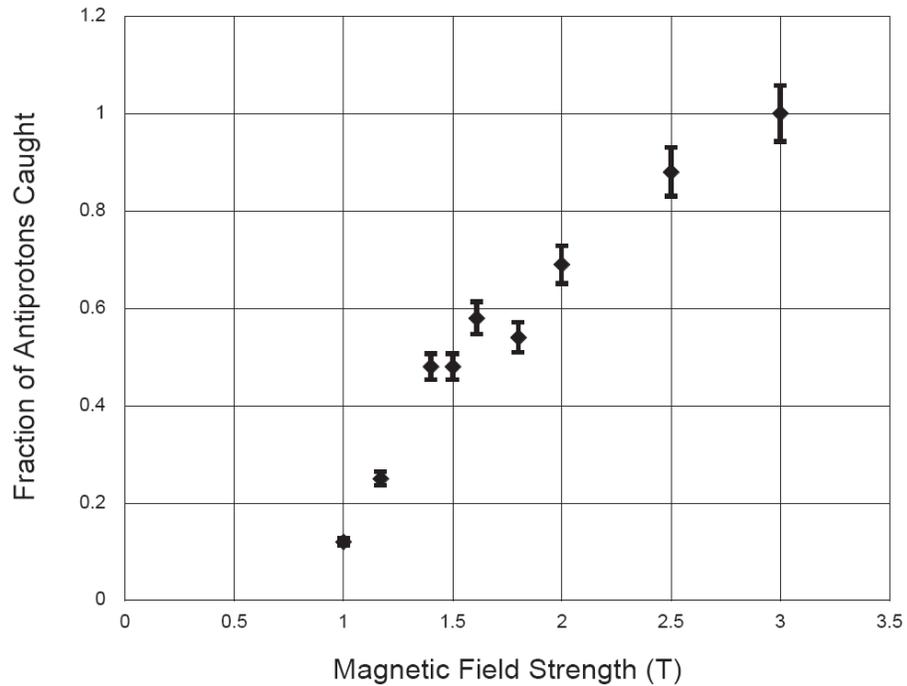

Fig. 2: Preliminary results for antiproton trapping efficiency versus magnetic field strength. The measurements are normalized to the value for 3 T field. A relative error of 6% is assumed for all the data points to represent the shot-to-shot variations of the measurements.

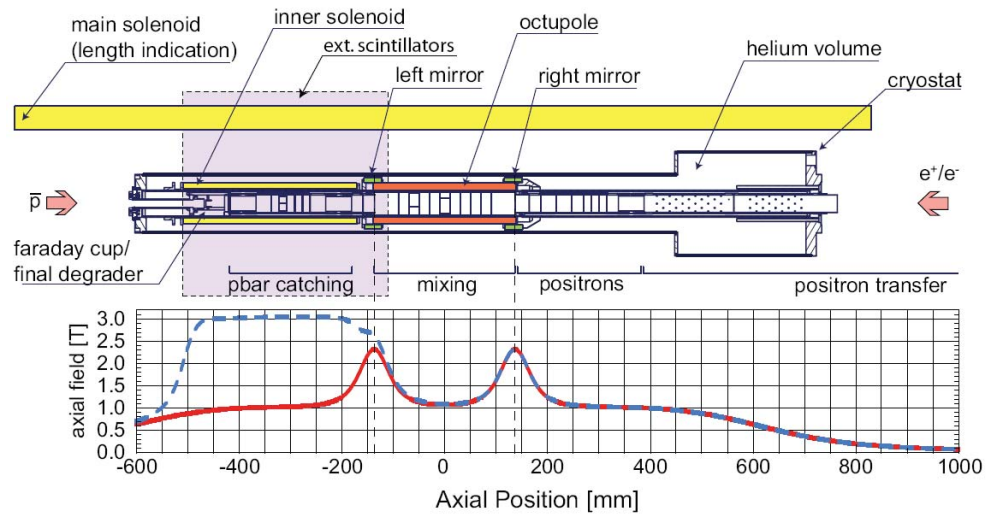

Fig. 3: Schematic view of the ALPHA antihydrogen trap. The graph shows the longitudinal magnetic field on axis due to the solenoids and mirror coils. The dashed curve is the field with the inner solenoid energized, whilst solid line is that without.

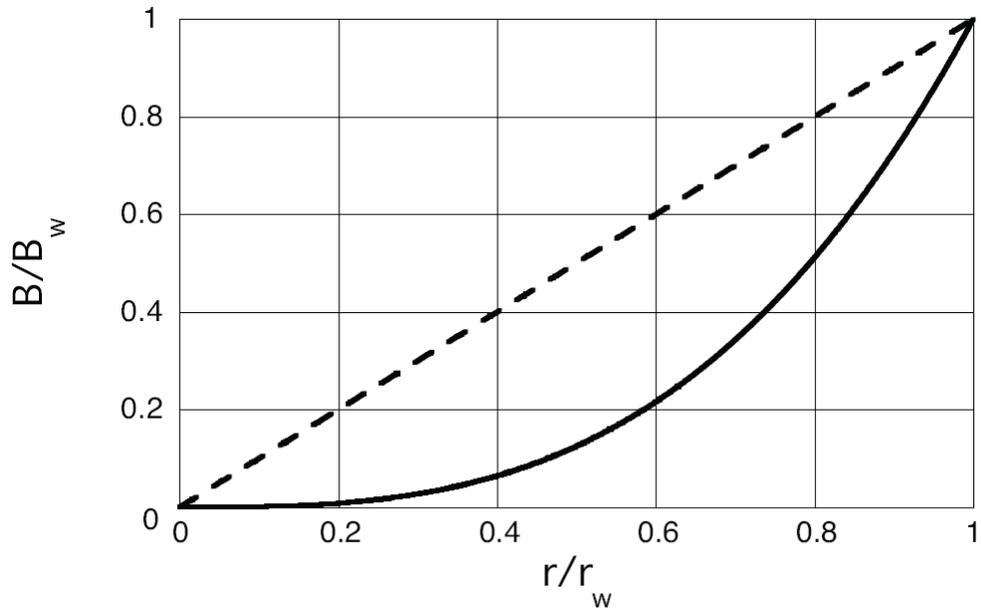

Fig. 4: Ideal radial profiles of the magnetic field for quadrupole (dashed line) and octupole (solid line) coil configurations. $B_w$ is the field at the inner wall of the Penning trap electrode, of radius $r_w$.

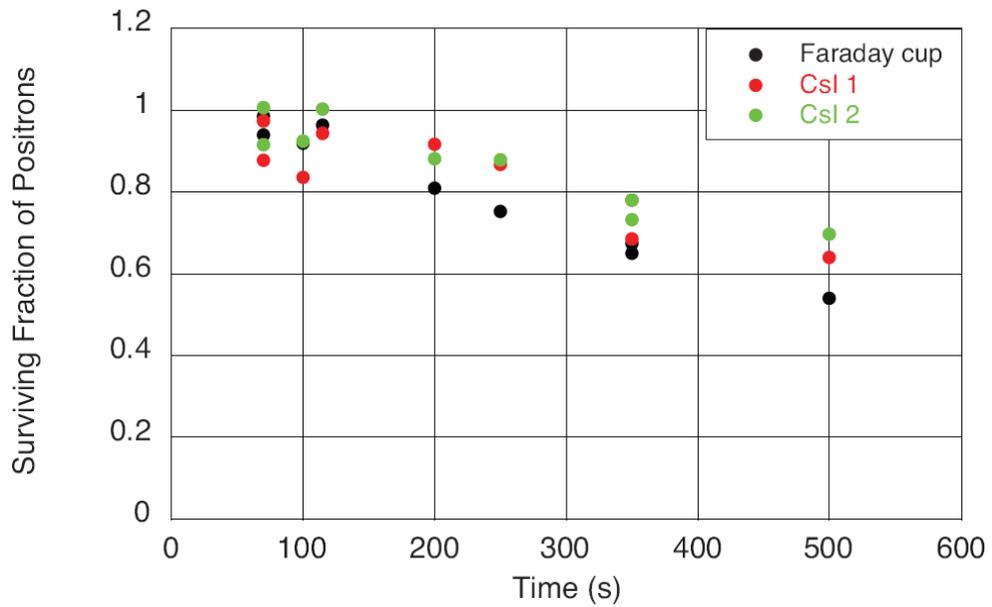

Fig. 5 : The ratio of the number of positrons stored in an octupole field to the number stored without the field is plotted versus holding time, as measured with a Faraday cup and with two CsI detectors.

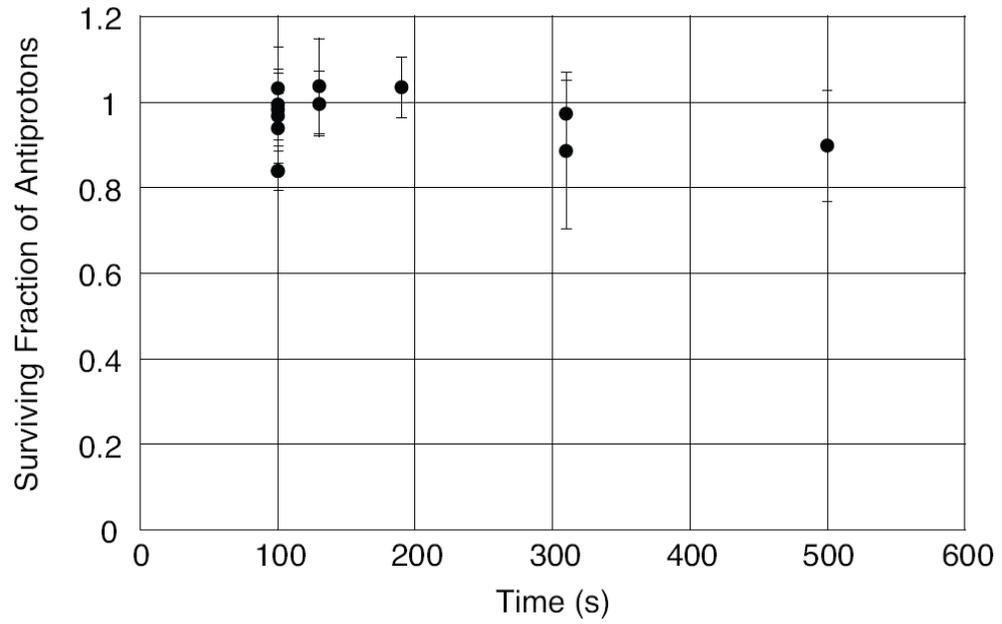

Fig. 6 : The ratio of the number of antiprotons stored in an octupole field to the number stored without the field is plotted versus holding time.

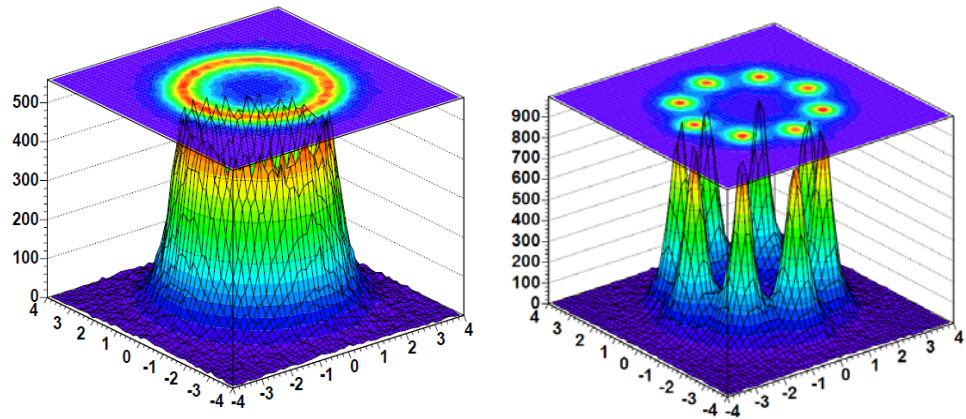

Fig. 7: Simulated distribution of reconstructed annihilation events on an electrode ring (4 cm diameter). [Left] Antihydrogen annihilations [Right] Antiproton loss to the wall, in the radial octupole field. The X and Y axes represent the vertex position in cm (the origin is the trap axis), whilst the z axis is the number of counts.

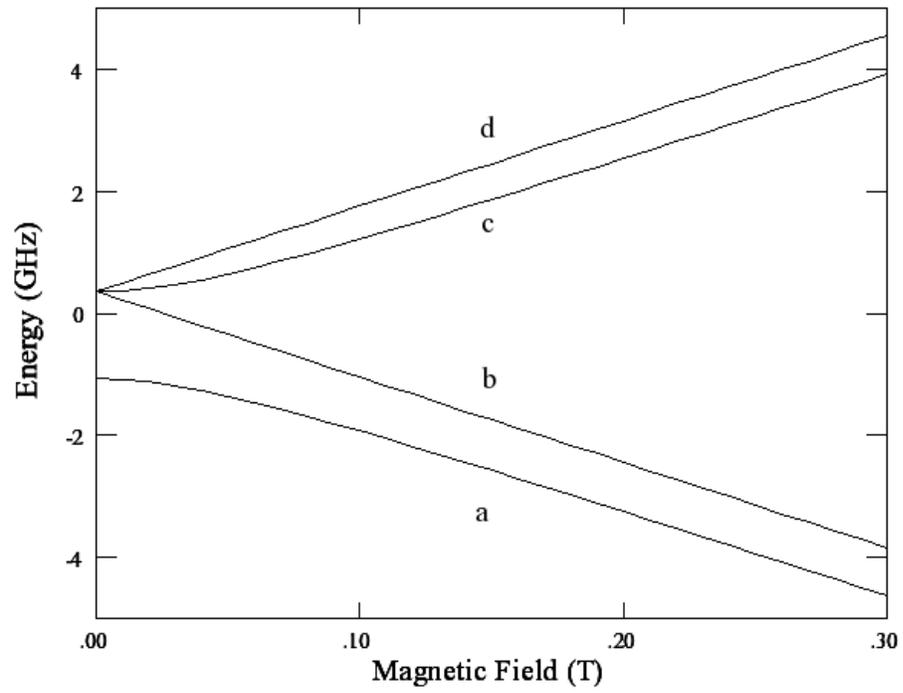

Fig 8: Breit-Rabi diagram for energy levels of (anti)hydrogen atoms in a magnetic field. See text for details.